\documentclass[preprint,showpacs,superscriptaddress]{revtex4-1}
\usepackage{graphicx}
\usepackage{amsmath}
\usepackage{amssymb}
\usepackage{color}

\begin{document}
\title{Traffic model by braking capability and response time}

\author{Hyun Keun Lee}
\affiliation{Department of Physics and Astronomy, Seoul National University,
Seoul 151-747, Korea}
\author{Jeenu Kim}
\affiliation{School of Computational Sciences, Korea Institute for Advanced Study,
Seoul 130-722, Korea}
\author{Youngho Kim}
\affiliation{The Korea Transport Institute, Goyang 411-701, Korea}
\author{Choong-Ki Lee}
\affiliation{School of Physics, Korea Institute for Advanced Study,
Seoul 130-722, Korea}
\date{\today}

\begin{abstract}
We propose a microscopic traffic model where the update velocity is determined by the deceleration capacity and response time. It is found that there is a class of collisions that cannot be distinguished by simply comparing the stop positions. The model generates the safe, comfortable, and efficient traffic flow in numerical simulations with the reasonable values of the parameters, and this is analytically supported. Our approach provides a new perspective in modeling the traffic-flow safety and the perturbing situations like lane change.
\end{abstract}


\maketitle

\begin{section}{introduction}
Modeling of traffic flow has been an intensive research topic for more than a half century in the engineering and science communities~\cite{LighthillPRS55,Pipes53,Chandler58,GazisOR59,HermanJORSJ62,Prigogine,Gipps81,NaSch91,KernerPRE93,BandoPRE95,TreiberPRE00}, of which results are summarized in the reviews~\cite{carev,BrackstoneTRF99,HelbingRMP01,NagataniRPG02}. In the progress of information technology, the various traffic models are required for the control and/or the automation of traffic flow. The needs is basically the credible modeling of safety and mobility, the two categorical
but conflicting goals in driving. Thus the natural driving behaviors have been modeled, for example, the more (less) acceleration for the larger (smaller) spacing.
However, this is usually based on
the trial-function approach,
as criticized in Ref~\cite{BrackstoneTRF99}.

There were a few seminal works that do not use trial function; one is Gipps (collision-avoidance) model~\cite{Gipps81,BrackstoneTRF99} and another one is Nagel-Schreckenberg (minimal collision-free) model~\cite{NaSch91,carev}. In Nagel-Schreckenberg model, any magnitude of deceleration is applied when required to prevent a collision. This means the deceleration capacity is actually unbounded (the so-called intelligent-braking-behavior suggested in~\cite{TreiberPRE00} also belongs to this case). Meanwhile, in Gipps model, a collision-avoidance in bounded deceleration capacity was suggested. We consider this approach is more physical, and thus we adopt it in the present work.

In this work, we examine how the collision can be understood in the physical constraints of the deceleration capacity and the response time. We propose a microscopic traffic model where the update velocity is determined in the safety criterion by the constraints. It is found that there is a class of collisions that cannot be identified by the usual safety criterion comparing the emergency-stop positions.
The resultant model generates in numerical test the practically appealing traffic flow of
safety, efficiency, and comfort with the reasonable parameter values, and this is analytically supported. Our model also provides a new perspective in modeling traffic-flow safety and the perturbing situations like lane change.
\end{section}

\begin{section}{modeling }
Since the safety and mobility are conflicting to each other,
a compromise between them is necessary.
A natural one is
the condition where driver can marginally avoid collision against the leader's full stop. Let $x_n(t)$ and $v_n(t)$ be the (front-end) position and velocity at time $t$, respectively, of vehicle $n$. The safety criterion is asking, in the presence of response time $\tau_n$, what is the marginal
$x_{n}(t+\tau_n)$ and $v_{n}(t+\tau_n)$ that does not result in a collision with the maximum deceleration $D_n$ from $t+\tau_n$, if the front vehicle at $x_{n+1}(t)$ and $v_{n+1}(t)$ begins to decelerate with its maximum deceleration $D_{n+1}$ from $t$ to stop. In short, this asks whether the worst situation is manageable in the physical constraint of braking capability and response time. Obviously, such a worst case may not happen. But it is necessary to check whether the follower can keep safe in that situation with its braking capacity and response time.

One of the safety criteria is shown in Fig.~\ref{match}(a), where $x_n(t+\tau_n)$ and $v_n(t+\tau_n)$ are adjusted so that the two trajectories become tangential as the two vehicles stop.
\begin{figure}
\includegraphics*[width=\columnwidth]{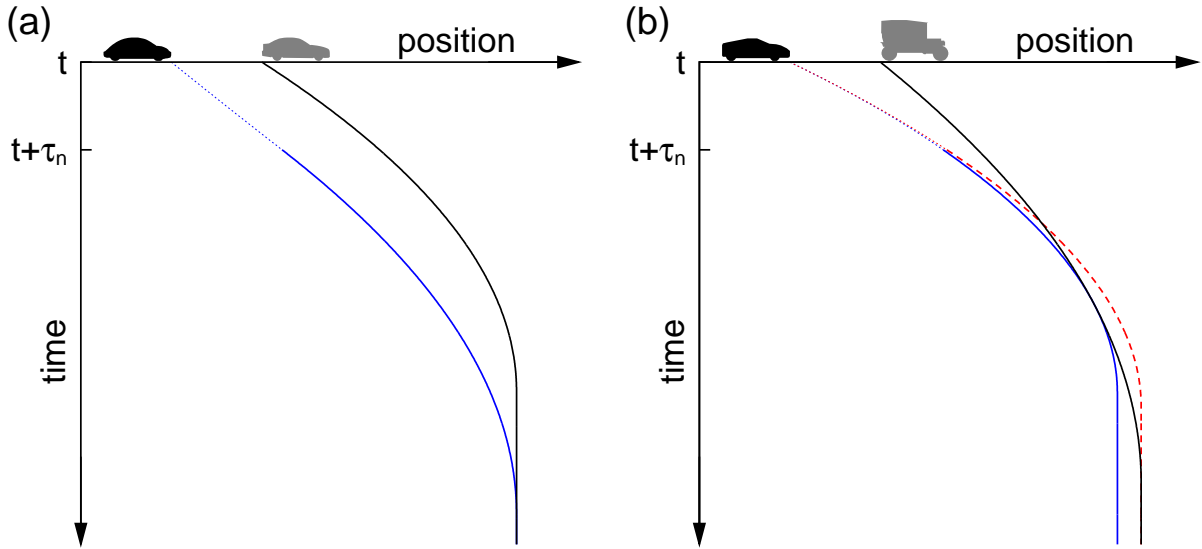}
\caption{(Color online) Two types of safety criterion in emergency when vehicles decelerate by their own braking capacities till stop. Follower's deceleration as the response to the emergency is delayed by response time $\tau_n$ (the dotted segment from $t$ and $t+\tau_n$ is not the part of emergency). The curvature of trajectory is given by the associated braking capacity. In criterion (a), the two stop positions are compared to tell a collision.
In criterion (b), the blue and black solid trajectories
meet though the follower's stop position does not exceed the leader's (this is possible only when the follower's braking capability is stronger than the leader's). Note this kind of collisions cannot be distinguished by comparing the stop positions (see the red dashed curve).}
\label{match}
\end{figure}
It basically compares the two vehicles' stop positions to tell a collision. This is same to that considered in Gipps model~\cite{Gipps81} believed so far to provide a safe enough dynamics.
Here, we point out that this safety criterion only is incomplete. This is because there is the other kind of collisions that cannot be discerned by comparing the stop positions, as follows.

The other kind is shown in Fig~\ref{match}(b), which is possible only when $D_n > D_{n+1}$. In this case, since the follower's trajectory is bent stronger than the leader's, the match of the stop positions (see the red-dashed curve)
necessarily brings about a collision before stop.
This collision is, however, not distinguished by simply comparing the stop positions.
It is thus necessary to reconsider the configuration at $t+\tau_n$. The follower's blue solid trajectory in Fig.~\ref{match}(b) is the alternative, which is adjusted to be tangential to the trajectory of the leader still in move. We remark that, even for $D_n > D_{n+1}$, there is a situation where the criterion of Fig.~\ref{match}(a) should still apply,
for example, if the follower is not so close to the leader at time $t$.

$x_n(t+\tau_n)$ and $v_n(t+\tau_n)$ are related by position-update rule. When the scheme of constant acceleration between responses is used, the position update reads
\begin{equation}
\label{xn}
x_n(t+\tau_n) = x_n(t) + \frac{\tau_n}{2} \left[ v_n(t) + v_n(t+\tau_n) \right].
\end{equation}
Considering this in the two tangential conditions explained above,
as the two marginal velocities at $t+\tau_n$,
one can obtain
\begin{equation}
\label{vsd}
\begin{split}
v_n^{\rm s}(t+\tau_n) &= -\frac{\tau_n D_n}{2}
+ \sqrt{ \left( \frac{\tau_n D_n}{2} \right)^2 + D_n \left(
	2 s_n(t)- \tau_n v_n(t) + \frac{v_{n+1}^2(t)}{D_{n+1}} \right) } , \\
v_n^{\rm d}(t+\tau_n) &= v_{n+1}(t) - \frac{\tau_n}{2} \left( D_n + D_{n+1} \right)
+ \sqrt{ \left( \frac{\tau_n \Delta D_n}{2} \right)^2 - \Delta D_n \left(
	2 s_n(t) + \tau_n \Delta v_n(t) \right) }~,
\end{split}
\end{equation}
where $\Delta D_n \equiv D_{n+1} - D_n$, $\Delta v_n(t) \equiv v_{n+1}(t) - v_n(t)$, and $s_n(t) = x_{n+1}(t) - x_n(t) - L_{n+1}$ for the leading vehicle's length $L_{n+1}$.

$v_n^{\rm s}(t+\tau_n)$ is enough to tell a collision when $D_n \le D_{n+1}$, while it is not when $D_n > D_{n+1}$.
Thus in the latter case, one of $v_n^{\rm s}(t+\tau_n)$ or $v_n^{\rm d}(t+\tau_n)$ should be selected depending on situations. Although Fig.~\ref{match}(b) shows a situation where $v_n^{\rm d}(t+\tau_n)$ should be selected, this is not always the case. If the follower is not so close to the leader, one can easily argue that $v_n^{\rm s}(t+\tau_n)$ is instead the proper choice. This way, considering a few conditions,
one knows the candidate of the update velocity is given by
\begin{equation}
\label{sel}
v_n^{\rm cand}(t+\tau_n) =
\begin{cases}
v_n^{\rm d}(t+\tau_n)	& \mbox{if~~} D_n > D_{n+1}, \,
			v_n^{\rm d}(t+\tau_n) \mbox{~is real}, \mbox{~and~}
			\tau_n + \frac{v_n^{\rm d}(t+\tau_n)}{D_n}
				< \frac{v_{n+1}(t)}{D_{n+1}}, \\
v_n^{\rm s}(t+\tau_n)	& \mbox{else if~} v_n^{\rm s}(t+\tau_n) \mbox{~is real}, \\
v_n(t) - \tau_n D_n^+	& \mbox{otherwise},
\end{cases}
\end{equation}
where the last case is introduced to cover such a situation allowing no physically meaningful $v_n^{\rm d}(t+\tau_n)$ and $v_n^{\rm s}(t+\tau_n)$.
For example, a careless cutting-in may bring it about.
This means there is no way to avoid a collision if the cutter brakes maximally till stops.
$D_n^+$ is introduced as an indicator of such an emergency that requires a deceleration larger than $D_n$, which is not possible by the definition of $D_n$. In order to simply cover that situation in the model, one may assign a value larger than $D_n$ to $D_n^+$.
We anticipate that the third case is crucial in modeling (un)tolerable perturbations.

$v_n^{\rm cand}(t+\tau_n)$ in Eq.~(\ref{sel}) is still the candidate velocity of the next step because its realizability from the current velocity is not taken into account yet.
The realizability is determined in the vehicular performance represented by the deceleration and acceleration capacities.
Thus when an acceleration capacity $A_n$ is additionally introduced,
the realizability corresponds to ``$-D_n \leq (v_n^{\rm cand}(t+\tau_n)-v_n(t)) / \tau_n \leq A_n$'', named as ``mechanical restriction''\cite{Lee04}. When a velocity change exceeding this range is required, only $-\tau_n D_n$ or $\tau_n A_n$ is possible by the definition of $D_n$ and $A_n$. Finally, considering the traffic regulation also, we arrive at
\begin{equation} \label{vn}
v_n(t+\tau_n) =
\max \{ v_n(t) - \tau_n D_n, \, 0,        \,
\min \{ v_n(t) + \tau_n A_n, \, v_{\max}, \, v_n^{\rm cand}(t+\tau_n) \} \},
\end{equation}
where zero stands for the directionality and $v_{\max}$ is a speed limit.
This gives the velocity update with Eqs.~(\ref{vsd}) and (\ref{sel}), and then position update is followed in Eq.~(\ref{xn}). The update rule is applied in parallel to all vehicles in system.
\end{section}

\begin{section}{manageability and potential collision}
Before analyzing our new model,
we discuss a few implications of Eq.~(4). The interest is in the case when
\begin{equation}
\label{mng}
v_n^{\rm cand}(t+\tau_n) \ge v_n(t+\tau_n).
\end{equation}
We below call a vehicle holding Eq.~(\ref{mng}) {\it manageable} at time $t$.
The inequality says the realization of $v_n^{\rm cand}(t+\tau_n)$ is possible in the braking capacity $D_n$. This again implies, if a deceleration is required for safety, it is realizable. Interestingly, the manageability at $t$ lasts thereafter unless perturbed later.
This is attributed to
the fact that $v_n^{\rm s,d}(t+\tau_n)$ are constructed in a way to keep safety against the leader's worst behavior; the consecutive maximal braking to stop (if already stopped, it is assumed not to move).
Thus once all the vehicles in a system are manageable, it lasts forever and the traffic flow remains collision-free,
as long as no perturbation is applied. A closed system composed of vehicles initially at rest is such an example.

As a
perturbation, one may consider the insertion of a vehicle into a gap between two vehicles. When the insertion takes place, it is reasonable to examine the manageability of the follower and that of the new comer. One may call it {\it manageable insertion} when the two vehicles are manageable at that instant.
We consider the notion of manageable insertion is crucial in modeling
on-ramp and/or lane-change. Also, a strategy to the dilemma zone by the traffic signal can be examined as considering the insertion of a standing object.
This way, the manageability
may shed a light
in designing and/or modeling the (un)tolerable traffic perturbations.

From the other perspective, the non-manageability [violation of Eq.~(\ref{mng})] can give a measure for safety indicating a {\it possible collision}.
The situation of non-manageability results in collision if the leader really applies the maximal brake to stop.
Thus the statistics on the non-manageability can be a reasonable measure for the traffic-flow safety. Note the flow including non-manageable configurations does not necessarily result in collision. Therefore, the flow without a collision can be regarded as
dangerous
in our approach.
We believe this viewpoint should be applied to the real traffic. It is necessary to discern a traffic flow with potential collisions so as to prevent a traffic accident in advance.

\end{section}

\begin{section}{numerical result and analytic support}
The present model is a consequence of the physical meaning of $D_n$ and $\tau_n$. We thus examine the flow property while varying them in numerical study. For the other model parameters, we use the typical values; $v_{\max} = 110$ ${\rm km/h}$, $A_n = 1.5$ ${\rm m/s}^2$, and $L_n = 7.5$ m~\cite{Lee04,Knospe00}. A system of randomly distributed vehicles initially at rest in the 100-km-long circular road is tested for various vehicular density $\rho$ (the total number of vehicles divided by the road length). For $D_n$, a random value out of the interval $(m_D-w_D, m_D+w_D)$ is assigned for various $m_D$ and fixed $w_D = 1.5~{\rm m/s}^2$ ($w_D$-value turns out not to change the results qualitatively). Below, we will use $\tau_n = \tau$ for all $n$ for a simplicity of the numerical implementation.

We observe that there emerge three kinds of steady state traffic flow depending on $m_D$, $\tau$, and $\rho$. Figure~\ref{phase}(a) shows the three flows with the position-velocity snapshots, where each dot represents a car.
\begin{figure}
\includegraphics*[width=\columnwidth]{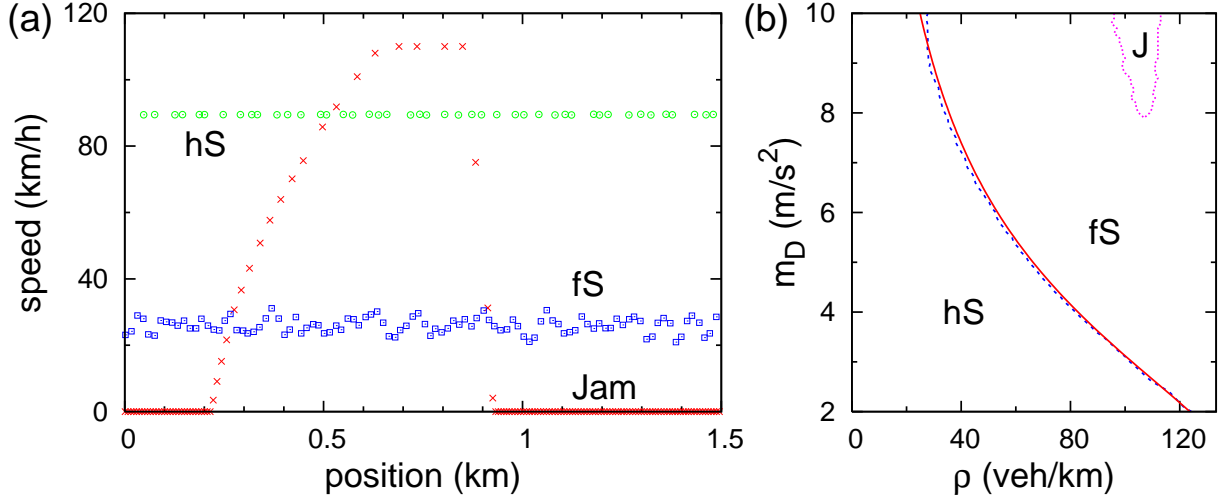}
\caption{(Color online) Three phases. (a) The location-speed snapshots for $\tau = 1$ sec. of hS flow ($\rho = 30~{\rm veh/km}$, $m_D = 5$ ${\rm m/s^2}$), fS flow ($\rho = 68$, $m_D = 5$), and jam phase ($\rho = 97$, $m_D = 10$). (b) The phase diagram in the parameter space $(\rho, m_D)$ for $\tau = 1$. Dotted (solid) lines are numerically (analytically) obtained phase boundaries.}
\label{phase}
\end{figure}
One of the snapshots (circles) shows an almost flat velocity profile, and we name it ``homogeneous steady'' (hS) flow. Another snapshot (squares) exhibits fluctuating velocities, named as ``fluctuating steady'' (fS) flow. Meanwhile, the last one (crosses) shows a traffic jam (J) where vehicles can hardly move. Figure~\ref{phase}(b) shows the phase diagram on $\rho$-$m_D$ plane for $\tau = 1$ sec., where the dotted (solid) curves are the numerically (analytically) obtained phase boundaries. We observe that the boundary between hS and fS is robust, while the region for J phase
depends on the initial configuration. In the following, we analytically argue that the observation above is the intrinsic feature of our model.

The hS flow is a homogeneous-velocity solution (HVS) of the model. Considering a constant velocity $v$ in Eq.~(\ref{sel}) for all $n$, one can find (see Appendix A)
\begin{equation}
\label{vh}
\frac{\rho^{-1} - L}{\tau^2\eta} = \begin{cases}
\frac{v}{\tau\eta}			& \mbox{for~~} \frac{v}{\tau\eta} \leq 1 , \\
f\left(\frac{v}{\tau\eta}\right)	& \mbox{for~~} \frac{v}{\tau\eta} > 1 \\
\end{cases}
\end{equation}
for $\eta^{-1} \equiv (m_D-w_D)^{-1}-(m_D+w_D)^{-1}$, where $f(z) = \left[ \int_{-1}^{1/z} \left( z - z^2 y / 2 \right) + \int_{1/z}^{1} 1 / 2 y \right] p(y) \, dy$ for $p(y)$ the probability distribution of the random variable $y$ standing for $\eta (1/D_{n+1}-1/D_n)$ [Eq.~(\ref{f4}) is the details of $f(z)$ for the $D_n$s we use].
As a general property indifferent to the statistics of $D_n$s, $f(z)$ is increasing and convex downward, and $f(z) \rightarrow z$ for $z \rightarrow 1$ while $f(z) \sim z^2$ for $z \gg 1$ [see the two limiting behaviors of the red curve (and also data points) in the upper-right part of Fig.~\ref{rv}(a)].

The solid curve in Fig.~\ref{rv}(a) is Eq.~(\ref{f4}) [a realization of
Eq.~(\ref{vh}) for the $D_n$s we use].
The numerical data for hS are perfectly on it above $v/\tau\eta = 1$. Therein, the circles and bars are, respectively, the velocity averages and fluctuations (see the latter is small enough to be covered in the data points for average).
\begin{figure}
\includegraphics*[width=\columnwidth]{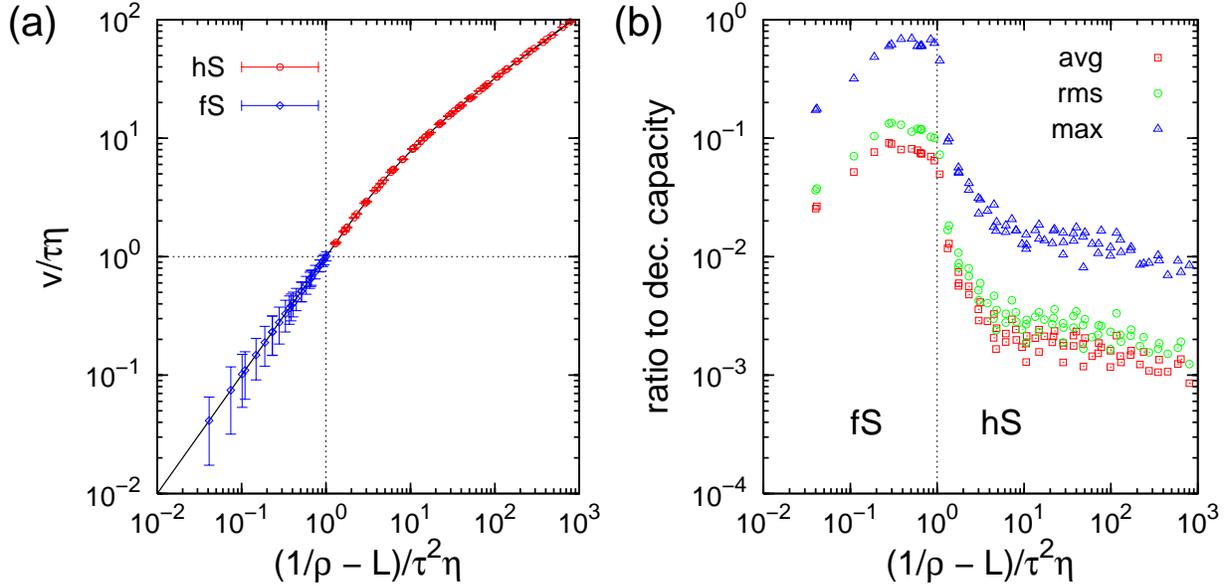}
\caption{(Color online) Density-velocity relation and deceleration. (a) Scaled density-velocity relation for $m_D = 5~{\rm m/s^2}$ and various $\tau = 1$, $1/2$, $1/4$, $1/8$, $1/16$ sec. The solid line is the analytic curve by Eq.~(\ref{vh}). Each point is the $10000$-sec. average of vehicles' velocities in steady state for a given density (all the points in Figs.~3 and 4 are obtained in the same way). The bar shows the fluctuation of the velocity. (b) Average, rms (root-mean-square) fluctuation, and maximum of the decelerations in the data giving (a) (each deceleration is measured in the ratio to the deceleration capacity).}
\label{rv}
\end{figure}
Interestingly, the average velocity of the numerical data for fS (diamonds) are also on the curve
in the other side,
even though there are considerable fluctuations
as indicated by the bars.
This suggests fS can also be understood with HVS. Below, we demonstrate that the dynamic property of HVS can explain this observations.

We performed the linear stability analysis~\cite{Strogatz94} on HVS and find this is linearly stable, including marginal stability, regardless of density and model parameters (see Appendix B). When $v/\tau\eta > 1$, there are at most two marginally stable modes out of the total $2 N$ stable eigenmodes, where $N$ is the number of vehicles. Otherwise with $v/\tau\eta \leq 1$, a half of the total modes are marginally stable. Thus when HVS is realized with $N \gg 1$ with $v/\tau\eta > 1$, it readily shows almost uniform velocity over the whole system while, with $v/\tau\eta \leq 1$, it may exhibit the fluctuations attributed to the macroscopic number of marginal modes. This dynamic property is consistent with the numerical observation on hS and fS shown in Fig.~\ref{phase}(a)~\cite{symmetry}.

It is worthy of noting that the stability boundary of $v/\tau\eta = 1$ is identical to that of the numerical phase boundary shown in Figs.~\ref{phase}(b) and \ref{rv}(a). When $v$ is replaced with $\rho$ using Eq.~(\ref{vh}), $(\rho^{-1}-L)/\tau^2\eta = 1$ is immediate. This is the phase boundary (solid curve) shown in Fig.~\ref{phase}(b), for $\tau = 1$. The macroscopic number of marginal modes in fS can explain the
observation of J in the fS-region (see Fig.~2(b)). Since the findings so far hold for any $\tau$ and statistics of $D_n$s, we conclude that hS and fS are the dynamic phases of our model. We finally emphasize the phase boundary condition of $v/\tau\eta = 1$ is same to the condition where at least one vehicle follows $v^{\rm d}$.
The flow established in the presence of such a vehicle is the very hS that is much more stable than fS. This indicates that $v^{\rm d}$ unrecognized in the earlier models plays a significant role in stabilizing the whole system.
\end{section}

\begin{section}{comfort and flux}
In the following, we examine our model generates a practically appealing traffic flow. If traffic flow is safe, one of the next concerns is the driving comfort, which is required for autonomous driving systems~\cite{vanArem06,Kesting08}. For this, we measure the decelerations each vehicles experience in the simulation for Fig.~\ref{rv}(a). Each deceleration is measured in the ratio to the deceleration capacity.
The results are shown in Fig.~\ref{rv}(b).
We obtain three statistical observables; average, fluctuation (root-mean-square), and maximum of the ratios. The average is approximately 0.1 and $0.001\sim 0.01$ in hS and fS flows, respectively, and the root-mean-square shows the similar values. The maximum is around $0.01\sim 0.1$ and $0.2\sim 0.7$, respectively.
We remark the deceleration strength is a characteristics of flow phase as observed, and thus driving comfort can be considerably improved by promoting hS with smaller $\tau$ (see Fig.~\ref{tau}).
\begin{figure}
\includegraphics*[width=\columnwidth]{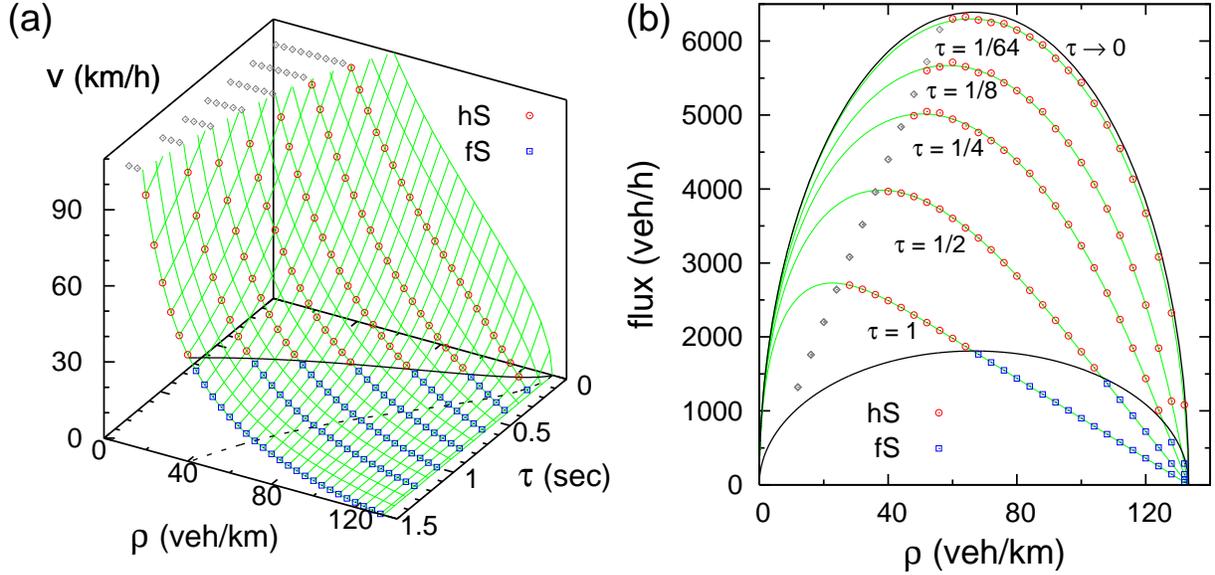}
\caption{(Color online) Effect of response time. (a) Average speed as a function of the density $\rho$ and the response time $\tau$ when $m_D = 5~{\rm m/s^2}$. The plateau in the upper left corner is speed limit $v_{\max}$. The solid curve separating the data point type is the phase boundary between hS and fS, and the dashed curve below is its projection onto the $\rho$-$\tau$ plane. (b) The projection of (a) onto the density-flux plane for $\tau = 1, 1/2, 1/4, 1/8$, and $1/64$ sec. The upper-bounding curve is obtained in the $\tau \rightarrow 0$ limit [see Eq.~(\ref{fluxlim})]. All curves in (a) and (b) are analytic results.}
\label{tau}
\end{figure}

The other practical interest is probably the flow efficiency, which can be represented by vehicular flux. For a homogeneous-velocity solution $v$, the flux is $\rho v$ by the hydrodynamic relation~\cite{HelbingRMP01}. Since $2 f(z) > z f^{\prime}(z)$ as a general property of $f(z)$ (Appendix A), the steady-state velocity $v$ increases as $\tau$ decreases. The velocity-increase for smaller $\tau$ is drawn in Fig.~\ref{tau}(a) along each constant-$\rho$ curves. This gives the flux-increase, as shown in Fig.~\ref{tau}(b). Since $f(z) \sim z^2$ for large $z$, the flux converges to
\begin{equation}
\label{fluxlim}
C \sqrt{\rho (1/L -\rho)}
\end{equation}
in the $\tau \rightarrow 0$ limit [see the upper-bounding solid curve in Fig.~\ref{tau}(b)], where $C$ is a constant by the statistics of $D_n$s [see Eq.~(\ref{Csquare}) for the detail]. We consider this result is also appealing because i) the flux for $\tau = 1$ sec. (the typical response time of drivers~\cite{Green00}) is comparable to empirical maximum value around $2500~{\rm veh/h}$~\cite{Hall86,Kerner98,Kim08}, ii) the flux enhancement is more sensitive for larger $\tau$, iii) the flux becomes considerable for $\tau$ of 0.1-sec.-order, and iv) all these are achieved in the manageable condition of Eq.~(\ref{mng}) guaranteeing safety. It is worthy of noting that the research field of autonomous driving system has already been treating the processing time down to 0.1 sec.~\cite{vanArem06,Kesting08}.
\end{section}

\begin{section}{final remark}
We finally remark that an extension of our model
in order to cover the other features of traffic flow (lane change, on-ramp flow, traffic signal, and so forth) is straightforward. This is because the problem is still a compromise between safety and mobility in consideration of the positions, velocities, and deceleration capacities of the related objects. Also, our model provides a new perspective to the study of traffic-flow safety through the interpretation of the non-manageable events and its statistics. Besides, we expect an autonomous driving system based on our model can be possible in the solid safety criterion and in the appealing traffic-flow quality of comfort and flux.
\end{section}

\begin{acknowledgments}
This research was supported by a grant (07-innovations in techniques A01) from the National Transportation Core Technology Program funded by Ministry of Land, Transport and Maritime Affairs of Korean government. This work was also supported by the grant of NRF of Korea (No. 2014R1A3A2069005).
\end{acknowledgments}

\begin{appendix}
\begin{section}{Homogeneous Solution}
Let a velocity $v$ be the homogeneous solution of our model.
Substituting
it
for all the velocities
in
Eq.~(\ref{vsd}),
we obtain the optimal spacing as
\begin{equation} \label{Sn}
S_n =
\begin{cases}
\tau v - \frac{1}{2} \left( \frac{1}{D_{n+1}} - \frac{1}{D_n} \right) v^2
& \mbox{for~~} \frac{1}{D_{n+1}} - \frac{1}{D_n} \leq \frac{\tau}{v} , \\
\frac{\tau^2}{2 \left( \frac{1}{D_{n+1}} - \frac{1}{D_n} \right)}
& \mbox{for~~} \frac{1}{D_{n+1}} - \frac{1}{D_n} >    \frac{\tau}{v} .
\end{cases}
\end{equation}
Then the average spacing is $S=\sum_{n=1}^N S_n / N$,
where $N$ is the total number of the vehicles.
The (global) vehicular density is simply given by
\begin{equation} \label{rhoV}
\rho = \frac{1}{S + L} ,
\end{equation}
and the average flux is
\begin{equation} \label{q}
q = \rho v .
\end{equation}

Equation~(\ref{Sn}) can be written as
\begin{equation} \label{svy}
s(v, \, y) =
\begin{cases}
s^{\rm s}(v, \, y) =
\tau v - \frac{1}{2} y v^2	& \mbox{for~~} y \leq \tau / v , \\
s^{\rm d}(v, \, y) =
\frac{\tau^2}{2 y}		& \mbox{for~~} y >    \tau / v ,
\end{cases}
\end{equation}
where $y$ stands for $\Delta d_n \equiv 1/D_{n+1} - 1/D_n$.
When $D_n$ is randomly assigned out of the interval $(m_D - w_D, \, m_D + w_D)$,
the range of $\Delta d_n$ is $(-1/\eta, \, 1/\eta)$ where
\begin{equation} \label{eta}
\frac{1}{\eta} \equiv \frac{1}{m_D-w_D} - \frac{1}{m_D+w_D} .
\end{equation}
Note that $v^{\rm d}$ (and accordingly $s^{\rm d}$) does not appear
when $v < \tau \eta$.

If there are sufficiently many vehicles
and their $D_n$'s are uncorrelated,
the average spacing can be obtained by the integral
\begin{equation} \label{SV}
S = \int_{-1/\eta}^{1/\eta} s(v, \, y) \, P_{\Delta d}(y) \, dy
= \int_{-1/\eta}^{\tau/v} \left( \tau v - \frac{v^2 y}{2} \right)
P_{\Delta d}(y) \, dy
+ \int_{\tau/v}^{1/\eta} \frac{\tau^2}{2 y} P_{\Delta d}(y) \, dy ,
\end{equation}
where $P_{\Delta d}(y)$ is the probability density function of
$\Delta d_n = 1/D_{n+1} - 1/D_n$.
Thus the average spacing of homogeneous solution with $v$
is completely determined by the distribution of $\Delta d_n$,
which can be obtained directly from the distribution of $D_n$.
Let us denote the probability density function of $D_n$ as $P_D(y)$.
Then $P_{d}(y)$ of $1/D_n$ is given by $P_{d}(y) = (1/y^2) P_D(1/y)$
and $P_{\Delta d}(y)$ is given by the convolution
$P_{\Delta d}(y)
= \int_{-\infty}^{\infty} P_{d}(x) \, P_{d}(x-y) \, dx$.
Note that $P_{\Delta d}(y)$ is an even function,
$P_{\Delta d}(-y) = P_{\Delta d}(y)$.

Let us introduce the dimensionless speed $z \equiv v / \tau \eta$
and the rescaled probability function $p(y)$ for $\eta (1/D_{n+1} - 1/D_n)$.
Then Eq.~(\ref{SV}) can be written as
\begin{equation} \label{SV3}
\frac{S}{\tau^2 \eta}
= \int_{-1}^{1/z} \left( z - \frac{z^2 y}{2}\right) p(y) \, dy
+ \int_{1/z}^{1} \frac{1}{2 y} p(y) \, dy .
\end{equation}
Since $p(y)$ has the normalization $\int_{-1}^{1} p(y) \, dy = 1$
and the symmetry property $p(-y) = p(y)$, we have
$\int_{-1}^{1/z} p(y) \, dy = 1 - \int_{1/z}^{1} p(y) \, dy$ and
$\int_{-1}^{1/z} y p(y) \, dy = - \int_{1/z}^{1} p(y) \, dy$.
Therefore we obtain
\begin{equation} \label{SV4}
\frac{S}{\tau^2 \eta} =
z \left[ 1 - I_0(z) \right] + \frac{z^2}{2} I_1(z) + \frac{1}{2} I_{-1}(z)
\equiv F(z) ,
\end{equation}
where
\begin{equation} \label{Ik}
I_k(z) \equiv \int_{1/z}^{1} y^k p(y) \, dy .
\end{equation}
This provides the scaling relation between $S$ (or $\rho$) and $v$,
\begin{equation} \label{sc1}
x \equiv \frac{S}{\tau^2 \eta}
= \frac{\rho^{-1} - L}{\tau^2 \eta}
= F\left(\frac{v}{\tau \eta}\right) = F(z).
\end{equation}


Now we find the general properties of $F(z)$.
Since $p(y) = 0$ for $|y| > 1$, we obtain $I_k(z) = 0$ for $z \leq 1$,
leading to $F(z) = z$ for $z \leq 1$,
or equivalently, $q = (1 - \rho L) / \tau$ for $\rho \geq 1/ (\tau^2 \eta + L)$.
Therefore we arrive at
\begin{equation} \label{Fz}
F(z) =
\begin{cases}
z	& {\rm for} \;\; z \leq 1 , \\
f(z)	& {\rm for} \;\; z  >   1 ,
\end{cases}
\end{equation}
where the function $f(z)$ is determined by the distribution $p(y)$.
Since $I_k(z = 1) = 0$, we obtain $\lim_{z \rightarrow 1^+} f(z) = 1 = F(1)$.
Thus $F(z)$ is a continuous function.


We have
$f(z) = z/2 + \left[ \int_0^{1/z} z + \int_{1/z}^{1} \left(
z^2 y / 2 + 1 / 2 y \right) \right] p(y) \, dy$
from $I_0(z) = 1/2 - \int_{0}^{1/z} p(y) \, dy$.
Since $\int_0^1 p(y) = 1/2$ and the integrand function
has the minimum $z$ and the maximum $(z^2+1)/2$ over $[0, \, 1]$,
the integral is bounded between $z/2$ and $(z^2+1)/4$.
Therefore we obtain
\begin{equation} \label{Fb}
z \leq F(z) \leq \frac{1}{4} (z+1)^2 .
\end{equation}
The lower bound $F(z) \geq z$ leads to $v \leq (\rho^{-1} - L) / \tau$,
which gives the upper bound of the average flux $q \leq (1 - \rho L) / \tau$.


For $z \leq 1$, we have $F'(z) = 1$.
Using
\begin{equation} \label{dI}
I'_k(z) = \frac{d}{dz} \int_{1/z}^{1} y^k p(y) \, dy
= \frac{1}{z^{k+2}} p\left( \frac{1}{z} \right) ,
\end{equation}
we obtain
$f'(z) = 1 - I_0(z) + z I_1(z)
= 1/2 + \left[ \int_{0}^{1/z} 1 + \int_{1/z}^{1} zy \right] p(y) \, dy$.
Since the integrand has the minimum $1$ and the maximum $z$ for $z > 1$,
the integral is bounded between $1/2$ and $z/2$.
Thus we have
\begin{equation} \label{dF}
1 \leq f'(z) \leq \frac{z+1}{2} .
\end{equation}
Since $F'(z) > 0$ for all $z > 0$, there exists
the inverse function $G = F^{-1}$ such that $G(F(z)) = G(x) = z$.
Therefore we obtain the inverse relation of (\ref{sc1}) as
\begin{equation} \label{Gx}
z = \frac{v}{\tau \eta}
= G\left( \frac{\rho^{-1} - L}{\tau^2 \eta} \right) = G(x) =
\begin{cases}
x	& {\rm for} \;\; x \leq 1 , \\
g(x)	& {\rm for} \;\; x  >   1 .
\end{cases}
\end{equation}
From (\ref{Fb}) and (\ref{dF}), we obtain $2\sqrt{x} - 1 \leq G(x) \leq x$.
Since $G'(x) = 1/F'(z)$, Eq.~(\ref{dF}) leads to
$1/\sqrt{x} \leq g'(x) \leq 1$.
From $dv/d\rho = - G'(x)/\tau \rho^2$,
we obtain $-1/\tau \rho^2 \leq dv/d\rho < 0$.


From (\ref{dF}) and (\ref{dI}), we obtain
\begin{equation} \label{d2F}
F''(z) = I_1(z) - I'_0(z) + z I'_1(z) = I_1(z) \geq 0 ,
\end{equation}
leading to $G''(x) = - F''(z)/(F'(z))^3 \leq 0$.
This implies the average flux $q$ is a non-concave function of $\rho$
because $d^2 q / d \rho^2 = G''(x) / \tau^3 \eta \rho^3 \leq 0$.


From (\ref{Gx}), $\partial v / \partial \tau$ at a fixed $\rho$ is given by
$\eta ( G(x) - 2 x G'(x) ) = - \eta (2 F(z) - z F'(z))/F'(z)$.
For $z \leq 1$, we have $2 F(z) - z F'(z) = z > 0$.
From (\ref{SV}) and (\ref{dF}), we obtain
$2 f(z) - z f'(z) = z \left( 1 - I_0(z) \right) + I_{-1}(z)
= z + z \int_{1/z}^{1} \left( 1/zy - 1 \right) p(y) \, dy \geq z $.
Thus we have
\begin{equation}
0 < \frac{2 F(z) - z F'(z)}{F'(z)} \leq z ,
\end{equation}
leading to $ - (\rho^{-1} - L) / \tau^2 \leq \partial v / \partial \tau < 0$.


There is another lower and upper bound of $F(z)$.
Using $I_0(\infty) = 1/2$, we have
$f(z) = z^2 I_1(\infty) / 2 + z / 2
+ \left[ \int_0^{1/z} ( z - z^2 y / 2 )
+ \int_{1/z}^1 (1 / 2 y) \right] p(y) \, dy $.
Since the integrand has the maximum $z$ and the minimum $1/2$,
the integral is bounded between $1/4$ and $z/2$.
Therefore we obtain
\begin{equation} \label{fb3}
\frac{z}{2} + \frac{1}{4}  \leq  f(z) - \frac{z^2}{2} I_1(\infty)  \leq  z .
\end{equation}
This leads to the asymptotic relation
\begin{equation} \label{fa}
F(z) \simeq \frac{z^2}{2} I_1(\infty) \quad \mbox{for~} z \gg 1,
\end{equation}
which implies
$q \simeq \sqrt{(2\eta/I_1(\infty)) \rho ( 1 - \rho L ) }$
for $\rho^{-1} - L \gg \tau^2 \eta$.


If $D_n$ has uniform distribution on
$( m_D - w_D, \, m_D + w_D ) \equiv (a, \, b)$,
then $P_{D}(y) = 1 / 2 w_D$, which leads to $P_{d}(y) = 1 / 2 w_D y^2$.
Thus we obtain
\begin{equation} \label{PD}
P_{\Delta d}(y) =
\begin{cases}
R(|y|)	& \mbox{for~~} |y| \leq 1/\eta , \\
0	& \mbox{for~~} |y| >    1/\eta ,
\end{cases}
\end{equation}
where
\begin{equation}
R(y) = \frac{ b - a - \frac{a}{1-ay} + \frac{b}{1+by} - \frac{2}{y}
	\ln \left[(1-ay)(1+by)\right] } {4 w_D^2 y^2} .
\end{equation}
After some algebra, we obtain
\begin{equation} \label{f4}
\begin{split}
f(z) &= z + \frac{1}{4 w_D^2} \left\{
\frac{2}{3} \left( 1 - z \right) \left( a b z + w_D^2 \right)
  - \frac{\eta^2 z^3}{3}
	\left[ \ln{ \left( 1 - \frac{a}{\eta z} \right) }
	     + \ln{ \left( 1 + \frac{b}{\eta z} \right) } \right]
\right. \\ & \left.
+ \frac{\eta z^2}{2}
	\left[ a \ln{\left( \frac{b}{a} - \frac{b}{\eta z} \right)}
	     - b \ln{\left( \frac{a}{b} + \frac{a}{\eta z} \right)} \right]
- \frac{1}{6 \eta}
	\left[ a^3 \ln{\left( \frac{bz}{a} - \frac{b}{\eta} \right)}
	     - b^3 \ln{\left( \frac{az}{b} + \frac{a}{\eta} \right)} \right]
\right\} .
\end{split}
\end{equation}


In the limit of $\tau \rightarrow 0$, we have
$s(v, y) = -\frac{1}{2} y v^2$ for $y \leq 0$ and $0$ for $y > 0$.
Then the average spacing is given by
\begin{equation}
S_{\tau \rightarrow 0} = \frac{v^2}{C^2} ,
\end{equation}
where
\begin{equation}
C^2 = \frac{2\eta}{\int_{0}^{1} y \, p(y) \, dy} .
\end{equation}
Then the average speed and flux are given by
\begin{equation} \label{qt0}
v_{\tau \rightarrow 0} = C \sqrt{\left( \frac{1}{\rho} - L \right)} ,\quad
q_{\tau \rightarrow 0} = C \sqrt{\rho \left( 1 - \rho L \right)}.
\end{equation}
For the uniform distribution of $D_n$ over $(m_D - w_D , m_D + w_D)$,
we obtain
\begin{equation}
\label{Csquare}
C^2 = \frac{2 w_D^2}{m_D \tanh^{-1} \left( \frac{w_D}{m_D} \right) - w_D} .
\end{equation}

\end{section}

\begin{section}{Linear Stability Analysis}
We investigate the linear stability
of the homogeneous solution with respect to small perturbations.
Assuming that the velocity and spacing are very close
to those of the homogeneous solution,
we can write down
\begin{equation} \label{us}
v_n(t) = v + u_n(t) , \quad s_n(t) = S_n + \sigma_n(t) ,
\end{equation}
where the optimal spacing $S_n$ is given by (\ref{Sn}).
By linearizing (3), we obtain
\begin{equation} \label{vsdt}
\begin{split}
v_n^{\rm s}(t+\tau) & = v
+ \frac{\sigma_n(t) - \frac{\tau}{2} u_n(t) + \frac{v}{D_{n+1}} u_{n+1}(t)}
{ \frac{\tau}{2} + \frac{v}{D_n} } , \\
v_n^{\rm d}(t+\tau) & = v + u_{n+1}(t)
+ \frac{D_n - D_{n+1}}{D_n + D_{n+1}}
\left[ \frac{2 \sigma_n(t)}{\tau} - u_n(t) + u_{n+1}(t) \right] ,
\end{split}
\end{equation}
up to the first order of $u_n$ and $\sigma_n$.
Meanwhile, from the integration scheme (1), we obtain
\begin{equation} \label{snt}
\sigma_n(t+\tau) = \sigma_n(t) + \frac{\tau}{2} \left[
u_{n+1}(t) + u_{n+1}(t+\tau) - u_n(t) - u_n(t+\tau) \right].
\end{equation}
The periodic boundary condition $x_{N+1} = x_1$
is applied for any quantity $x$.

We introduce the new variable
\begin{equation} \label{psi}
\psi_n(t) \equiv \frac{\sigma_n(t)}{\tau}
		+ \frac{u_n(t)}{2} - \frac{u_{n+1}(t)}{2} .
\end{equation}
From (\ref{snt}), we obtain
\begin{equation} \label{psit}
\psi_n(t+\tau) = \psi_n(t) - u_n(t) + u_{n+1}(t) .
\end{equation}
Then we can rewrite (\ref{vsdt}) as
\begin{equation} \label{usdt}
\begin{split}
v_n^{\rm s}(t+\tau) - v & = \frac{\psi_n(t)}{\alpha_n} - \frac{u_n(t)}{\alpha_n}
			+ \frac{\alpha_{n+1}}{\alpha_n} \, u_{n+1}(t) , \\
v_n^{\rm d}(t+\tau) - v & = \beta_n \, \psi_n(t) - \beta_n \, u_n(t)
			+ ( 1 + \beta_n) \, u_{n+1}(t) ,
\end{split}
\end{equation}
where
\begin{equation} \label{alpha}
\alpha_n \equiv \frac{1}{2} + \frac{v}{\tau D_n} , \quad
\beta_n \equiv 2 \, \frac{D_n - D_{n+1}}{D_n + D_{n+1}} .
\end{equation}
Note that $\alpha_n > 1/2$ and $0 < 1/\alpha_n < 2$.
The selection criterion $1/D_{n+1} - 1/D_n > \tau / v$ for $v_n^{\rm d}$
is equivalent to $\alpha_{n+1} - \alpha_n > 1$.
Now we combine (\ref{usdt}) into
\begin{equation} \label{ut}
u_n(t+\tau) = \gamma_n \, \psi_n(t)
		- \gamma_n \, u_n(t) + \theta_n \, u_{n+1}(t) ,
\end{equation}
where
\begin{equation} \label{gamma}
\begin{split}
\gamma_n = \gamma_n^{\rm s} , \quad	\theta_n = \theta_n^{\rm s}
& \quad \mbox{for~~} \alpha_{n+1} - \alpha_n \leq 1 , \\
\gamma_n = \gamma_n^{\rm d} , \quad	\theta_n = \theta_n^{\rm d}
& \quad \mbox{for~~} \alpha_{n+1} - \alpha_n   >  1 ,
\end{split}
\end{equation}
and
\begin{equation} \label{gtsd}
\begin{split}
\gamma_n^{\rm s} = \frac{1}{\alpha_n} , \quad &
\theta_n^{\rm s} = \frac{\alpha_{n+1}}{\alpha_n} ,\\
\gamma_n^{\rm d} = \beta_n , \quad &
\theta_n^{\rm d} = 1 + \beta_n .
\end{split}
\end{equation}
Here we have assumed that the selection of $v_n^{\rm s}$ or $v_n^{\rm d}$
is not changed by $u_n$.
Note that $\gamma_n$ and $\theta_n$ are continuous functions
of $\alpha_n$ and $\alpha_{n+1}$.


By introducing the vector notation
$\mathbf{f}(t) \equiv ( \psi_1(t), \cdots, \psi_N(t) )$,
$\mathbf{u}(t) \equiv ( u_1(t), \cdots, u_N(t) )$, and
$\mathbf{x}(t) \equiv ( \mathbf{f}(t), \, \mathbf{u}(t) )$,
we can combine (\ref{psit}) and (\ref{ut}) into a Jacobian matrix equation
\begin{equation} \label{xt}
\mathbf{x}(t+\tau) =
\begin{pmatrix}
\mathbf{f}(t+\tau) \\ \mathbf{u}(t+\tau)
\end{pmatrix}
=
\begin{pmatrix}
			\mathbf{I}_N & \mathbf{B} \\
			\mathbf{C}   & \mathbf{T}
\end{pmatrix}
\begin{pmatrix}
			\mathbf{f}(t) \\ \mathbf{u}(t)
\end{pmatrix}
= \mathbf{J} \, \mathbf{x}(t) ,
\end{equation}
where $\mathbf{I}_N$ is the $N \times N$ identity matrix
and the lower-left submatrix $\mathbf{C}$ is a diagonal matrix
with $G_{nn} = \gamma_n$.
The upper-right submatrix $\mathbf{B}$ and
the lower-right submatrix $\mathbf{T}$ are given by
\begin{equation} \label{BT}
\mathbf{B} =
\begin{pmatrix}
	-1     & 1      & 0      & \cdots & 0      \\
	0      & -1     & 1      & \cdots & 0      \\
	0      & 0      & -1     & \cdots & 0      \\
	\vdots & \vdots & \vdots & \ddots & \vdots \\
	1      & 0      & 0      & \cdots & -1
\end{pmatrix}
\quad
\mathbf{T} =
\begin{pmatrix}
	-\gamma_1 & \theta_1  & 0         & \cdots & 0      \\
	0         & -\gamma_2 & \theta_2  & \cdots & 0      \\
	0         & 0         & -\gamma_3 & \cdots & 0      \\
	\vdots    & \vdots    & \vdots    & \ddots & \vdots \\
	\theta_N  & 0         & 0         & \cdots & -\gamma_N
\end{pmatrix}
\end{equation}

The long-time behavior of the perturbation amplitude is determined by
the largest magnitude among the eigenvalues.
The eigenvalues are given by the zeros of the characteristic polynomial
\begin{equation}
h(\lambda) \equiv
\det{\left( \mathbf{J} - \lambda \mathbf{I}_{2N} \right)}
= \det{ \begin{pmatrix} \mathbf{A} & \mathbf{B} \\
			\mathbf{C} & \mathbf{D} \end{pmatrix} } = 0 ,
\end{equation}
where $\mathbf{A} = (1-\lambda) \mathbf{I}_N$
and $\mathbf{D} = \mathbf{T} - \lambda \mathbf{I}_N$.
Using the properties of the determinant,
we obtain
\begin{equation} \label{h}
h(\lambda)
= \lambda^N \prod_{n=1}^N \mu_n(\lambda) - \prod_{n=1}^N \omega_n(\lambda)
= 0 ,
\end{equation}
where
\begin{equation}
\mu_n(\lambda) \equiv \lambda - 1 + \gamma_n , \quad
\omega_n(\lambda) \equiv \theta_n \left( \lambda - 1 \right) + \gamma_n .
\end{equation}

For $\lambda = 1$, we have $\mu_n(1) = \omega_n(1) = \gamma_n$,
leading to $h(1) = 0$.
Therefore at least one eigenvalue is exactly $1$.
From (\ref{h}), we have
$h'(1) = (\prod_{n=1}^N \gamma_n) \sum_{n=1}^N ( 1 + (1-\theta_n)/\gamma_n)$.
Since the summand is zero for $v_n^{\rm d}$ and positive for $v_n^{\rm s}$,
we obtain $h'(1) > 0$.
Thus the eigenvalue $1$ is unique without degeneracy.
On the other hand, for $\lambda = -1$, we have
$\omega_n(-1)/\mu_n(-1) = (2 \alpha_{n+1} - 1 )/(2 \alpha_n - 1 )$,
irrespective of whether $v_n^{\rm sel}$ is $v_n^{\rm s}$ or $v_n^{\rm d}$,
leading to $\prod_{n=1}^N (\omega_n(-1)/\mu_n(-1)) = 1$.
Thus we have $h(-1) = ( 1 - (-1)^N ) \prod_{n=1}^N (2 - \gamma_n)$,
which is zero for even $N$ and negative for odd $N$.
Thus $-1$ is an eigenvalue if and only if $N$ is even.
For even $N$, we obtain
$h'(-1) = (-1)^N (\prod_{n=1}^N (2-\gamma_n))
( -N + \sum_{n=1}^N ( 1/(2-\gamma_n/\theta_n) - 1/(2-\gamma_n) ) )$.
The sum is zero when $v_n^{\rm sel} = v_n^{\rm s}$ for all $n$.
Since $1/(2-\gamma_n^{\rm d}/\theta_n^{\rm d}) - 1/(2-\gamma_n^{\rm d})
< 1/(2-\gamma_n^{\rm s}/\theta_n^{\rm s}) - 1/(2-\gamma_n^{\rm s})$
for $\alpha_{n+1} - \alpha_n > 1$,
the sum is not positive.
Thus we obtain $h'(-1) \neq 0$ and $-1$ is also a unique eigenvalue.


Moreover, $\pm 1$ are the absolute bounds of real eigenvalues.
For $\lambda > 1$, we have $\mu_n(\lambda) > 0$ and $\omega_n(\lambda) > 0$.
Similarly, we have $\mu_n(\lambda) < 0$ and $\omega_n(\lambda) < 0$
for $\lambda < -1$ since $\gamma_n - 2 \theta_n < 0$.
For $\lambda$ to be a solution of (\ref{h}), it is necessary to satisfy
$|\prod_{n=1}^N \omega_n(\lambda)/\mu_n(\lambda)| = |\lambda|^N$.
Meanwhile, it can be easily shown that
\begin{equation} \label{ps}
\prod_{n=1}^N \left|
\frac{\omega_n^{\rm s}(\lambda)}{\mu_n^{\rm s}(\lambda)} \right| = 1 ,
\end{equation}
due to the transitivity
$\omega_n^{\rm s}(\lambda) = \theta_n^{\rm s} \mu_{n+1}^{\rm s}(\lambda)$
and $\prod_{n=1}^N \theta_n^{\rm s} = 1$.
For $|\lambda| > 1$, we obtain
\begin{equation} \label{asd}
  \left|\frac{\omega_n^{\rm d}(\lambda)}{\mu_n^{\rm d}(\lambda)}\right|
< \left|\frac{\omega_n^{\rm s}(\lambda)}{\mu_n^{\rm s}(\lambda)}\right|
\quad \mbox{for~~} \alpha_{n+1} - \alpha_n > 1 ,
\end{equation}
by using
\begin{equation} \label{nsd}
\left| \omega_n^{\rm s}(\lambda) \, \mu_n^{\rm d}(\lambda) \right| -
\left| \omega_n^{\rm d}(\lambda) \, \mu_n^{\rm s}(\lambda) \right| =
\frac{ ( \alpha_{n+1} - \alpha_n ) ( \alpha_{n+1} - \alpha_n - 1 ) }
	{ \alpha_n ( \alpha_{n+1} + \alpha_n - 1 ) }
\left( |\lambda|^2 - 1 \right) .
\end{equation}
Therefore we obtain
$\prod_{n=1}^N |\omega_n(\lambda)/\mu_n(\lambda)|
\leq 1 < |\lambda|^N$ for $|\lambda| > 1$.
Thus there is no real eigenvalues in the range $\lambda > 1$ or $\lambda < -1$.

The equality (\ref{ps}) and the inequality (\ref{asd}) holds
even for the complex $\lambda = r e^{i\phi}$
with the magnitude $r > 1$ and angle $\phi$ ($0 \leq \phi < 2\pi$).
After some algebra, we have
\begin{equation} \label{nsdc}
\begin{split}
& \left| \omega_n^{\rm s}(r e^{i\phi}) \, \mu_n^{\rm d}(r e^{i\phi}) \right|^2
- \left| \omega_n^{\rm d}(r e^{i\phi}) \, \mu_n^{\rm s}(r e^{i\phi}) \right|^2
\\ & =
\frac{ ( \alpha_{n+1} - \alpha_n ) [ ( \alpha_{n+1} - \alpha_n )^2 - 1 ] }
	{ \alpha_n^2 ( \alpha_{n+1} + \alpha_n - 1 ) }
\, 4 \left( 1 - \cos^{2}\phi \right)
+ \left(a_0 + a_1 \cos\phi \right) \left(r^2 - 1 \right)
			+ a_2 \left( r^2 - 1 \right)^2
\\ & \equiv M_n(r, \, \cos\phi) ,
\end{split}
\end{equation}
where
$a_0$, $a_1$, and $a_2$ are constants
to be determined by $\alpha_n$ and $\alpha_{n+1}$.
Due to (\ref{asd}), we have $M_n(r, \, \pm 1) > 0$ for $r > 1$,
which corresponds to $\lambda = \pm r$ ($\phi = 0$ or $\pi$).
Since $M_n(r, \, \cos\phi)$ is a quadratic function of $\cos\phi$
with the negative coefficient on the quadratic term.
Therefore $M_n(r > 1, \, \cos\phi)$ is positive over the entire range of $\phi$.
This leads to
\begin{equation} \label{asdc}
  \left|\frac{\omega_n^{\rm d}(r e^{i\phi})}{\mu_n^{\rm d}(r e^{i\phi})}\right|
< \left|\frac{\omega_n^{\rm s}(r e^{i\phi})}{\mu_n^{\rm s}(r e^{i\phi})}\right|
\quad \mbox{for~~} \alpha_{n+1} - \alpha_n > 1 ,
\end{equation}
for any angle $\phi$ if $r > 1$.
Thus we can conclude that
$\prod_{n=1}^N |\omega_n(r e^{i\phi})/\mu_n(r e^{i\phi})|
\leq 1 < r^N$ for $r > 1$.
Therefore there is no complex eigenvalues in the region $|\lambda| > 1$.


If $v^{\rm d}$ does not appear at all
($\rho > 1/(\tau^2 \eta + L)$),
we have $\gamma_n = 1/\alpha_n$
and $\theta_n = \alpha_{n+1} / \alpha_n$ for all $n$.
Then we have $\mu_n(\lambda) = \lambda - 1 + 1/\alpha_n$ and
$\omega_n(\lambda)
= ( \alpha_{n+1} / \alpha_n ) ( \lambda - 1 ) + 1/\alpha_n
= \alpha_{n+1} \mu_{n+1}(\lambda) / \alpha_n $,
leading to $\prod_{n=1}^N \omega_n(\lambda) = \prod_{n=1}^N \mu_n(\lambda)$.
Thus we obtain
\begin{equation}
\left( \lambda^N - 1 \right)
\prod_{n=1}^N \left( \lambda - 1 + \frac{1}{\alpha_n} \right) = 0 ,
\end{equation}
which leads to $\lambda = e^{2 \pi i n/N}$ or $\lambda = 1 - 1/\alpha_n$
($n = 1, \, \cdots , \, N$).
Since $| 1 - 1/\alpha_n | < 1$, we obtain $|\lambda|_{\max} = 1$.
Thus a $v^{\rm s}$-only homogeneous flow is {\it marginally stable}.
Note that $N$ eigenvalues among the total $2N$ eigenvalues have $|\lambda| = 1$.


On the other hand, if $v^{\rm d}$ is selected for at least one vehicle,
the situation changes drastically.
For $\lambda = e^{i\phi}$, we obtain
\begin{equation} \label{nsd1}
\left| \omega_n^{\rm s}(e^{i\phi}) \mu_n^{\rm d}(e^{i\phi}) \right|^2 -
\left| \omega_n^{\rm d}(e^{i\phi}) \mu_n^{\rm s}(e^{i\phi}) \right|^2 =
\frac{ ( \alpha_{n+1} - \alpha_n ) [ ( \alpha_{n+1} - \alpha_n )^2 - 1 ] }
	{ \alpha_n^2 ( \alpha_{n+1} + \alpha_n - 1 ) }
\, 4 \left( 1 - \cos^{2}\phi \right) ,
\end{equation}
by substituting $r = 1$ in Eq.~(\ref{nsdc}).
Since $-1 < \cos\phi < 1$ for any non-real $\lambda$
along the unit circle on the complex plane,
we have
\begin{equation} \label{asd1}
  \left|\frac{\omega_n^{\rm d}(e^{i\phi})}{\mu_n^{\rm d}(e^{i\phi})}\right|
< \left|\frac{\omega_n^{\rm s}(e^{i\phi})}{\mu_n^{\rm s}(e^{i\phi})}\right|
\quad \mbox{for~~} \alpha_{n+1} - \alpha_n > 1 ,
\end{equation}
unless $\phi$ is an integer multiple of $\pi$.
Thus we have $\prod_{n=1}^N |\omega_n(e^{i\phi})/\mu_n(e^{i\phi})| < 1$
for all non-real $\lambda = e^{i\phi}$
if at least one $v_n^{\rm d}$ is selected.
Therefore, all the eigenvalues except for $1$ (and $-1$ for even $N$)
have the magnitude less than $1$.
Consequently the $v^{\rm d}$-mixed flow is much more stable
than the $v^{\rm s}$-only flow.

\end{section}

\end{appendix}


\begin{thebibliography}{99}
\bibitem{symmetry} Whether a marginal property in linearized system will be maintained in the original nonlinear system is nontrivial. In fact, however, fS data reveals the period of $N\tau$	 (not shown here). This strongly suggests that fS phase is the combination of the periodic orbits deformed from the periodic marginal modes in the linearized system~\cite{Strogatz94}.
\end{thebibliography}
\end{document}